\documentclass[prl,twocolumn,floatfix,showpacs,superscriptaddress,reprint,citeautoscript]{revtex4-2}
\usepackage{color}
\usepackage{epsfig}
\usepackage{graphicx}
\usepackage{amsmath}
\usepackage{booktabs}
\usepackage{multirow}
\usepackage[table,xcdraw]{xcolor}
\usepackage[colorlinks, linkcolor=blue, citecolor=blue, urlcolor=blue]{hyper ref}

\begin{document}

\title{Distinguishing thermal fluctuations from polaron formation in halide perovskites}
\author{Bai-Qing Zhao}
\affiliation{School of Materials Science and Engineering, Northwestern Polytechnical University, Xi’an 710072, China}
\affiliation{Materials Department, University of California, Santa Barbara, CA 93106-5050, USA}

\author{Xuan-Yan Chen}
\affiliation{Sustainable Energy and Environment Thrust, Function Hub, Hong Kong University of Science and Technology (Guangzhou), Guangzhou 511458, China}

\author{Chuan-Nan Li}
\affiliation{Materials Department, University of California, Santa Barbara, CA 93106-5050, USA}

\author{Jinshan Li}
\email{ljsh@nwpu.edu.cn}
\affiliation{School of Materials Science and Engineering, Northwestern Polytechnical University, Xi’an 710072, China}

\author{Chris G. Van de Walle}
\email{vandewalle@mrl.ucsb.edu}
\affiliation{Materials Department, University of California, Santa Barbara, CA 93106-5050, USA}

\author{Xie Zhang}
\email{xie.zhang@nwpu.edu.cn}
\affiliation{School of Materials Science and Engineering, Northwestern Polytechnical University, Xi’an 710072, China}

\begin{abstract}
Recent angle-resolved photoelectron spectroscopy (ARPES) measurements of the hole effective mass in CsPbBr$_3$ revealed an enhancement of $\sim$50\% compared to the bare mass computed from first principles for CsPbBr$_3$ at $T = 0$~K.
This large enhancement was interpreted as evidence of polaron formation.
Employing accurate finite-temperature first-principles calculations, we show that the calculated hole effective mass of CsPbBr$_3$ at $T = 300$~K can explain experimental results without invoking polarons.
Thermal fluctuations are particularly strong in halide perovskites compared to conventional semiconductors such as Si and GaAs, and cannot be ignored when comparing with experiment.
We not only resolve the debate on polaron formation in halide perovskites, but also demonstrate the general importance of including thermal fluctuations in first-principles calculations for strongly anharmonic materials.
\end{abstract}

\maketitle

Lattice vibrations at finite temperatures mediate thermal fluctuations in atomic and electronic structures, which further impact the material properties~\cite{zhang_profiling_2024}.
While lattice vibrations are always present, they are usually not taken into account in first-principles calculations, even for the basic electronic band structure.
The reasons for this neglect are twofold. 
First, since the main focus is typically on material properties at room temperature, temperature effects are assumed to be minor.
Second, the density functional theory (DFT)~\cite{kohn_self_1965} that underlies the majority of first-principles calculations is strictly speaking only valid at $T = 0$~K, and incorporating temperature effects is challenging.
While remarkable progress has been made in theoretical computations of phonons (quasiparticles for lattice vibrations)~\cite{kresse_ab_1995,gonze_dynamical_1997,togo_first_2015,baroni_phonons_2001} 
and electron-phonon coupling~\cite{giustino_electron_2017,monserrat_electronphonon_2018} and of their role in band-structure renormalization~\cite{monserrat_anharmonic_2013,zacharias_theory_2020}, these calculations are often prohibitively expensive for routine use.
Pragmatically, for many materials lattice vibrations can indeed be disregarded while computing band structures and derived properties; however, there are important exceptions, especially when dealing with strongly anharmonic materials.

Halide perovskites, which have emerged as highly efficient materials for optoelectronics, are strongly anharmonic materials with a soft lattice~\cite{chouhan_synthesis_2020}.
Their excellent performance has often been attributed to the formation of polarons~\cite{miyata_large_2017,schlipf_carrier_2018}.
Evidence of polaron formation in halide perovskites is commonly based on a comparison of experimental and theoretical effective masses derived from the band structure.
Puppin \textit{et al.}~\cite{puppin_evidence_2020} performed angle-resolved photoelectron spectroscopy (ARPES) measurements for the valence-band dispersion of CsPbBr$_3$ perovskite, which revealed an enhancement of 50\% for the hole effective mass ($0.26\pm0.02$ $m_0$) compared to the value computed with hybrid-functional DFT (0.17~$m_0$).
The enhancement was interpreted as evidence for the formation of large polarons in CsPbBr$_3$.
This attribution was subsequently questioned by Sajedi \textit{et al.}~\cite{sajedi_is_2022}, who performed independent ARPES measurements and $GW$ calculations.
The mass computed by Sajedi \textit{et al.} (0.226~$m_0$) was even greater than their new experimental result ($0.203\pm0.016$~$m_0$), indicating an absence of polaron formation.
Recently, however, Rieger \textit{et al.}~\cite{rieger2023surface} reported an ARPES-measured hole effective mass of 0.3~$m_0$ for CsPbBr$_3$, greater than all of the theoretically computed hole masses.
In short, while the experimentally measured effective masses exhibit some spread, there may indeed be an enhancement compared to the first-principles results.
However, since the conclusion about polaron formation hinges on an experimentally observed enhancement of the effective mass, the key question is: what is the correct reference band structure of \textit{polaron-free} CsPbBr$_3$ that can serve as a reference to compare with the experimental results?

Rigorous computation of the electronic band structure of CsPbBr$_3$ is non-trivial for the following reasons.
First, as discussed above and also acknowledged in Ref.~\onlinecite{puppin_evidence_2020}, thermal fluctuations may also cause mass renormalization;
Puppin \textit{et al.}~\cite{puppin_evidence_2020} concluded that the impact of thermal fluctuations on effective masses is negligible, but provided only a rough estimation. 
Second, ``first-principles'' or ``DFT calculations'' are highly imprecise labels that cover a wide range of approaches with different levels of accuracy, depending, for instance, on 
the exchange-correlation functional or on specific parameters that can critically impact the final electronic band structure.
We note that neither the first-principles calculations with hybrid functionals by Puppin \textit{et al.}~\cite{puppin_evidence_2020} nor the $GW$ calculations by Sajedi \textit{et al.}~\cite{sajedi_is_2022} were able to reproduce the experimental band gap of CsPbBr$_3$.
This raises concerns about the derived effective masses, since the latter are correlated with the band gap, as shown by $k \cdot p$ theory~\cite{yu_fundamentals_2010}.

In this Letter, we rigorously investigate the electronic band structure of CsPbBr$_3$ from first principles, employing the Heyd-Scuseria-Ernzerhof (HSE)~\cite{heydHybrid2003} hybrid functional with spin-orbit coupling (SOC) included and taking into account the thermal lattice fluctuations at room temperature.
The reliability of this methodology has been experimentally verified in the computation of an accurate electronic band structure for CsPbI$_3$~\cite{zhao_engineering_2023}.
For CsPbBr$_3$, we also obtain a band gap in good agreement with experiment (2.37 eV)~\cite{li2021single}.
We find that, after taking into account thermal fluctuations at room temperature, and without introducing a hole in the valence band, our calculated hole effective mass agrees very well with the experimental results of Puppin \textit{et al.}~\cite{puppin_evidence_2020}.
Hence, referenced to a correct band structure, there is no evidence of a significant enhancement in the experimentally measured hole effective mass, and consequently no basis for concluding that polaron formation occurs in halide perovskites.

\begin{figure}[h]	
\includegraphics[width=8.6cm]{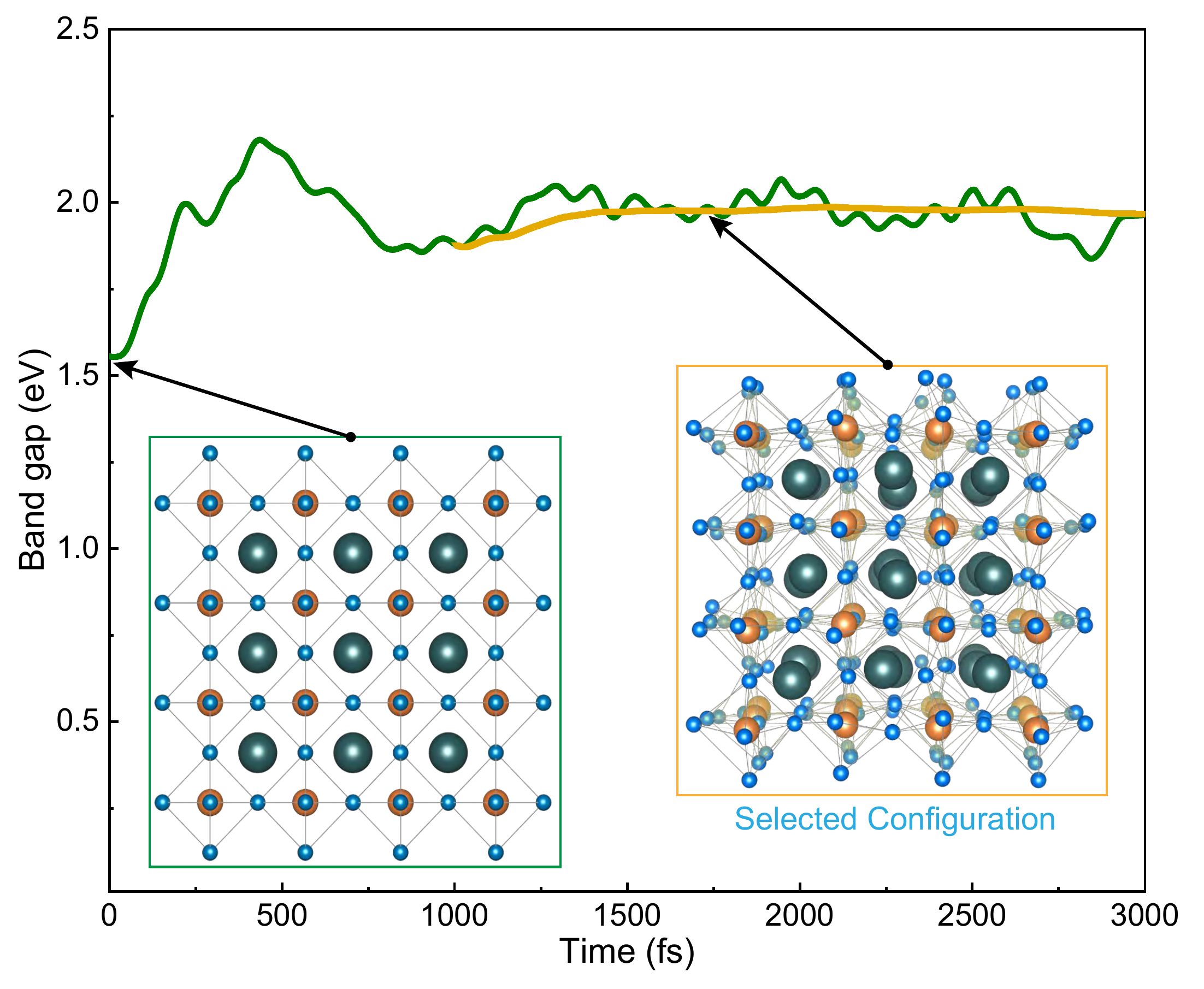}
\caption{
Evolution of the band gap of CsPbBr$_3$ as a function of time in an AIMD simulation.
The yellow line shows the accumulated average value of the band gap  (starting from 1000 fs). 
The insets show the atomic structures of the ideal cubic lattice at 0~K and distorted lattice at 300 K for CsPbBr$_3$.
}
\label{fig:MD}
\end{figure}

To obtain an accurate band structure for CsPbBr$_3$, a proper description of its atomic structure at room temperature is required.
We therefore perform \textit{ab initio} molecular dynamics (AIMD) simulations at 300~K using the canonical ensemble ($NVT$) with the Nos\'{e}-Hoover thermostat~\cite{noseUnified1984,hooverCanonical1985} as implemented in the Vienna \textit{ab initio} simulation package ({\sc vasp})~\cite{kresse_efficient_1996}. 
The initial structure is a $3 \times 3 \times 3$ supercell of the ideal cubic perovskite phase (lattice constant: 5.87~{\AA}~\cite{zheng_balanced_2018}) containing 135 atoms (see inset of Fig.~\ref{fig:MD}). 
We use the Perdew-Burke-Ernzerhof (PBE) functional~\cite{perdewGeneralized1996} (which has been demonstrated to accurately describe structural properties~\cite{chen_crystal-liquid_2023}) to simulate the structural evolution and equilibration; 
to evaluate the precise electronic band structure we will subsequently apply the HSE functional in conjunction with SOC.
The time step of the AIMD simulations is set to 1 fs.

Figure~\ref{fig:MD} shows the evolution of the band gap as a function of time.
Clearly, the band gap experiences a pronounced increase (on the order of 0.5 eV) in the first 1000 fs, and afterwards it gradually equilibrates with small fluctuations (see the yellow line in Fig.~\ref{fig:MD} for the accumulated average after 1000 fs).
This already implies that thermal lattice fluctuations significantly modulate the electronic structure of CsPbBr$_3$. 
Consistent with previous studies~\cite{zhao_engineering_2023} we find that after reaching thermodynamic equilibrium the electronic band structure remains fairly stable, though the atomic structure is still dynamically fluctuating. 
Hence, we select a representative pseudocubic atomic configuration (see inset of Fig.~\ref{fig:MD}) to evaluate its accurate electronic structure at room temperature.

\begin{figure*}[ht]	
\includegraphics[width=180mm]{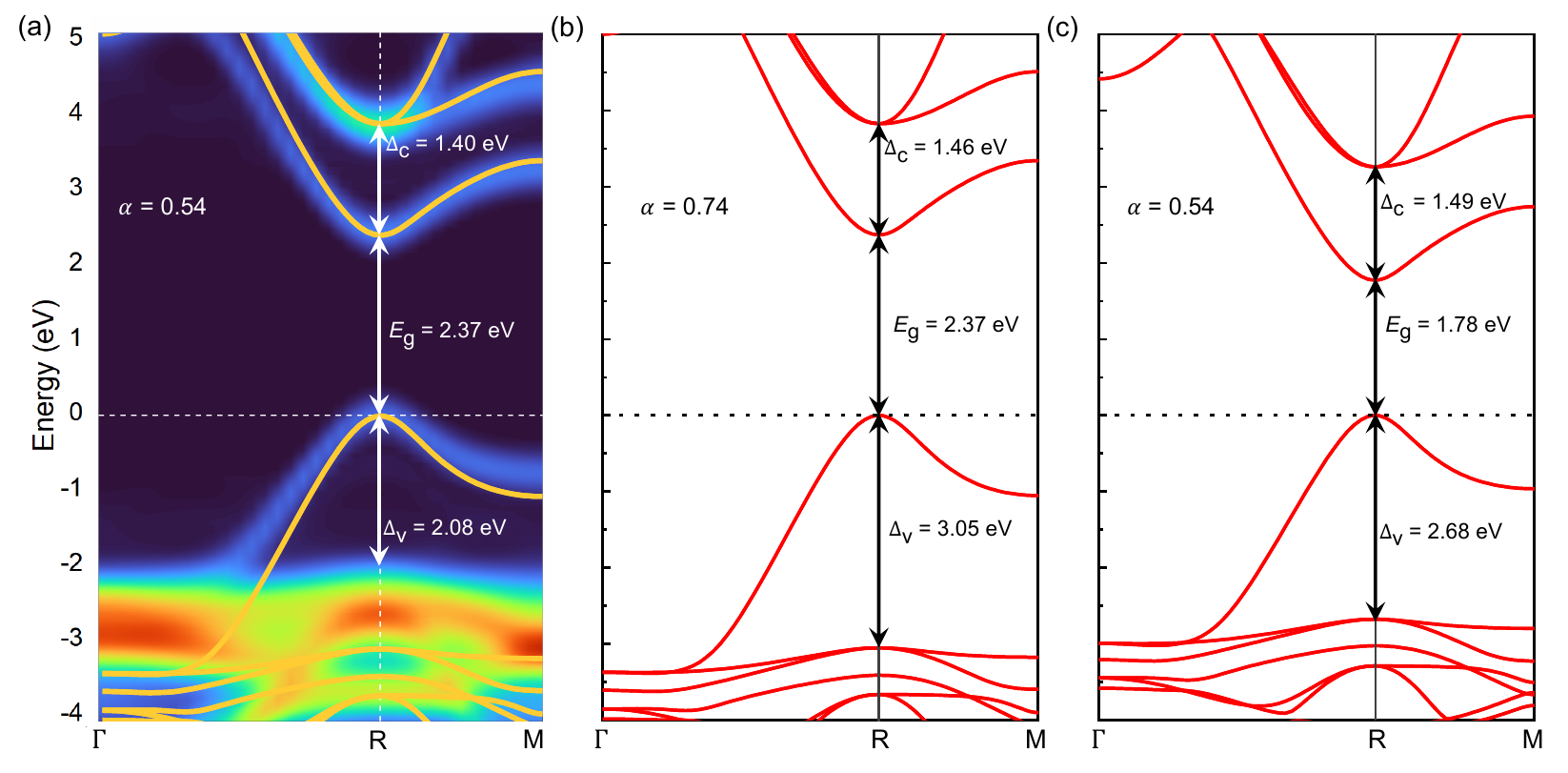}
\caption{
(a) Unfolded band structure of CsPbBr$_3$ with the representative AIMD configuration at room temperature.
(b,c) Band structure of cubic CsPbBr$_3$ computed from the HSE-SOC scheme with different fractions of nonlocal Fock exchange ($\alpha$).
The orange lines in (a) depict the computed band structure for ideal cubic CsPbBr$_3$ (at 0~K).
}
\label{fig:EBS}
\end{figure*}

One technical issue in utilizing the HSE functional is the determination of a suitable mixing parameter ($\alpha$), i.e., the fraction of nonlocal Fock exchange to include in the first-principles calculations.
We use the experimental room-temperature band gap (2.37~eV~\cite{li2021single}) as a benchmark criterion to determine $\alpha$, which leads to $\alpha = 0.54$.
The colormap in Fig.~\ref{fig:EBS}(a) shows the calculated band structure for the selected AIMD configuration in Fig.~\ref{fig:MD} by Brillouin-zone unfolding~\cite{shen_band_2017,wang_vaspkit_2021} onto the cubic symmetry.
The following results and discussion focus on the cubic phase of CsPbBr$_3$; we also performed calculations and analyses for the orthorhombic phase (see the Supplemental Material~\footnote{See Supplemental Material at [URL] for the results for the orthorhombic phase of CsPbBr$_3$ and AIMD simulations for CsPbBr$_3$ with an extra hole.}), which reveal similar trends.

To elucidate the importance of thermal lattice fluctuations, we also calculate the electronic band structure for ideal cubic CsPbBr$_3$ using the HSE-SOC scheme~\cite{zhangMinimizing2021} for comparison.
If we follow the same benchmark criterion, we actually need a much larger $\alpha$ of 0.74 to reproduce the room-temperature experimental band gap.
The band structure for ideal cubic CsPbBr$_3$ obtained in this fashion is depicted in Fig.~\ref{fig:EBS}(b) and also included in Fig.~\ref{fig:EBS}(a) (orange line) for comparison.

One can see that even though the band gaps are identical in the two cases, the band dispersions and thus the effective masses are different.
In addition, the intra-valence-band transition energy ($\Delta_v$) exhibits a very large difference of around 1.0 eV between the two cases.
$\Delta_v$ is a very special feature in the renormalized band structure of halide perovskites, and this reduction has been quantitatively confirmed by supercontinuum transient-absorption experiments~\cite{zhao_engineering_2023}.
As shown in Table~\ref{tab:mixing}, after taking into account the thermal lattice fluctuations both the electron and hole effective masses are significantly enhanced (57\% for $m_e$ and 79\% for $m_h$)  as compared to those for the ideal cubic structure computed with $\alpha = 0.74$.
The hole effective mass obtained for the AIMD structure with thermal fluctuations at room temperature included is 0.265~$m_0$, very close to the experimental value ($0.26\pm0.02$ $m_0$) in Ref.~\cite{puppin_evidence_2020}.

To exclude the impact of $\alpha$, we also compute the band structure for ideal cubic CsPbBr$_3$ with $\alpha=0.54$.
As shown in Fig.~\ref{fig:EBS}(c), we obtain a band gap  of 1.78 eV, which is actually very close to the value (1.74 eV) calculated with the many-body $GW$ approach in Ref.~\cite{sajedi_is_2022}.
This implicitly suggests that $\alpha$ = 0.54 is a more reasonable choice, but the use of correct atomic structure is key; $GW$ itself does not guarantee reliable band structure.
Using $\alpha=0.54$ on the ideal cubic structure not only significantly underestimates the true experimental gap (1.78 vs. 2.37 eV), but also significantly overestimates $\Delta_v$ (2.68 vs. 2.08 eV) and underestimates the effective masses (Table~\ref{tab:mixing}), demonstrating that the impact of thermal fluctuations is clearly very strong and the observed effects cannot be attributed to the choice of $\alpha$.

\begin{table}[h!]
\caption{Comparison of calculated band gaps and electron and hole effective masses ($m_e$ and $m_h$) with different structures and different fractions of nonlocal Fock exchange ($\alpha$) in the HSE hybrid functional.}
\label{tab:mixing}
\renewcommand{\arraystretch}{1.2}
\begin{ruledtabular}
\begin{tabular}{lcccc}
\multicolumn{2}{c}{cubic phase} & 300 K ($\alpha$=0.54) & 0 K ($\alpha$=0.74) & 0 K ($\alpha$=0.54) \\ \hline
\multicolumn{2}{c}{$E_{\rm g}$(eV)}       & 2.37   & 2.37   & 1.78   \\ \hline
\multirow{3}{*}{$m_e$ ($m_0$)} & R-M & 0.290 & 0.172 & 0.157 \\ 
                           & R-$\Gamma$ & 0.251 & 0.172 & 0.157 \\ 
                           & Ave. & 0.263 & 0.172 & 0.157 \\ \hline
\multirow{3}{*}{$m_h$ ($m_0$)} & R-M & 0.264 & 0.148 & 0.141  \\ 
                           & R-$\Gamma$ & 0.265 & 0.149 & 0.142  \\ 
                           & Ave. & 0.265 & 0.149 & 0.142 \\ 
\end{tabular}
\end{ruledtabular}
\end{table}

\begin{figure}[h]	
\includegraphics[width=8.5cm]{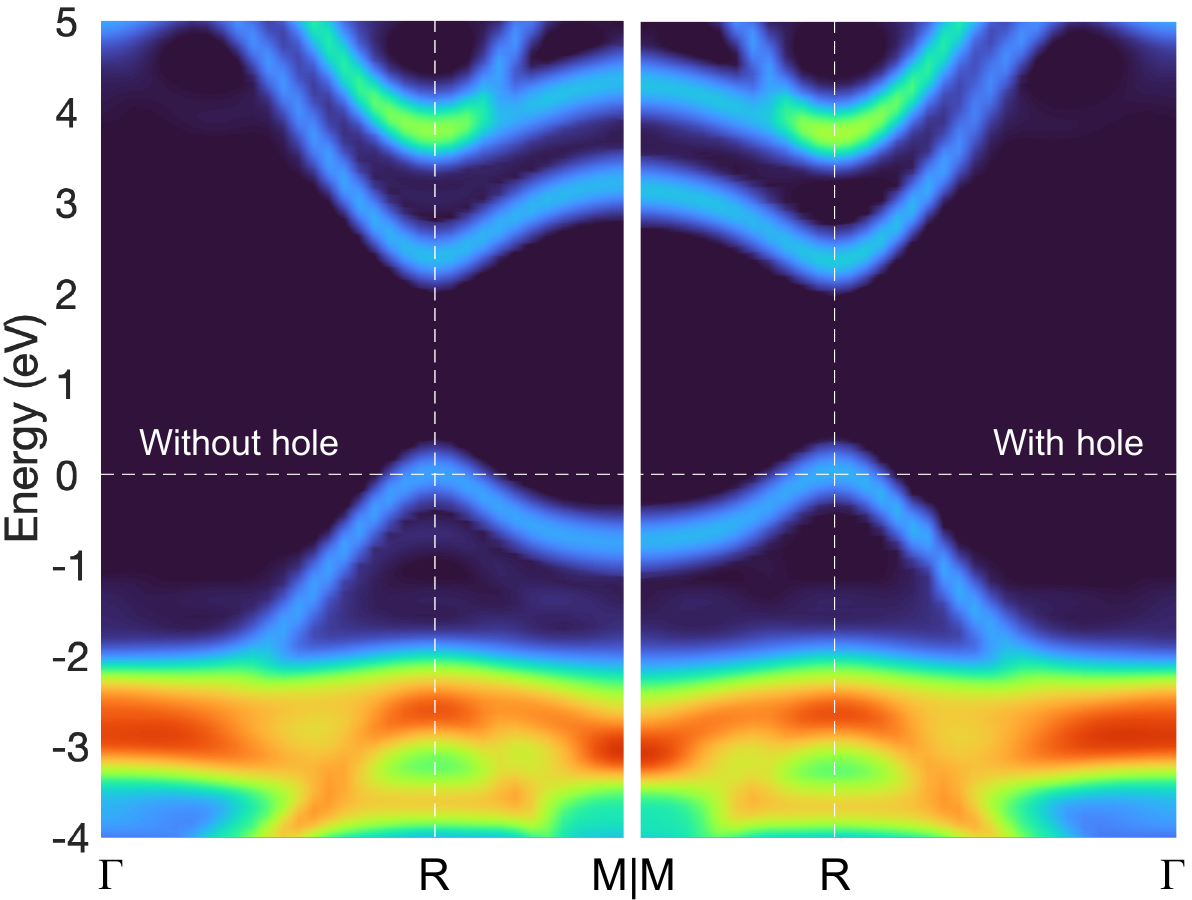}
\caption{
Unfolded band structure of CsPbBr$_3$ without and with an extra hole.}
\label{fig:EFF}
\end{figure}

We emphasize that the thermally induced mass enhancement occurs  \textit{in the absence of introducing a hole in the valence band}. In order to investigate whether the presence of such holes in the ARPES measurement would lead to any additional effects, we explicitly introduce an extra hole in the CsPbBr$_3$ supercell.
We again perform AIMD simulations at room temperature (see the Supplemental Material~\cite{Note1}) to extract a relevant atomic configuration for further calculation of its electronic band structure.
We employ the HSE-SOC scheme (with $\alpha = 0.54$) to compute the unfolded band structure for the equilibrated CsPbBr$_3$ supercell with an extra hole included at room temperature as shown in Fig.~\ref{fig:EFF}, in comparison with the electronic band structure of  \textit{polaron-free}  CsPbBr$_3$.
Two interesting observations can be made.

First, no localized gap states are formed in the band structure of CsPbBr$_3$ with an extra hole introduced at room temperature, which implies that small polarons are not formed in this material.
Second, the addition of an extra hole has a negligible impact on the electronic band structure of CsPbBr$_3$ (see Fig.~\ref{fig:EFF}).
We also explicitly checked whether hole localization can occur and did not find any evidence for polaron stability.
This analysis therefore does not produce any evidence for polaron formation.
We note that the enhancement of effective masses observed here has also clear distinctions from that caused by large polaron formation, since it appears in the absence of an extra charge.

\begin{figure}[ht]	
\includegraphics[width=86mm]{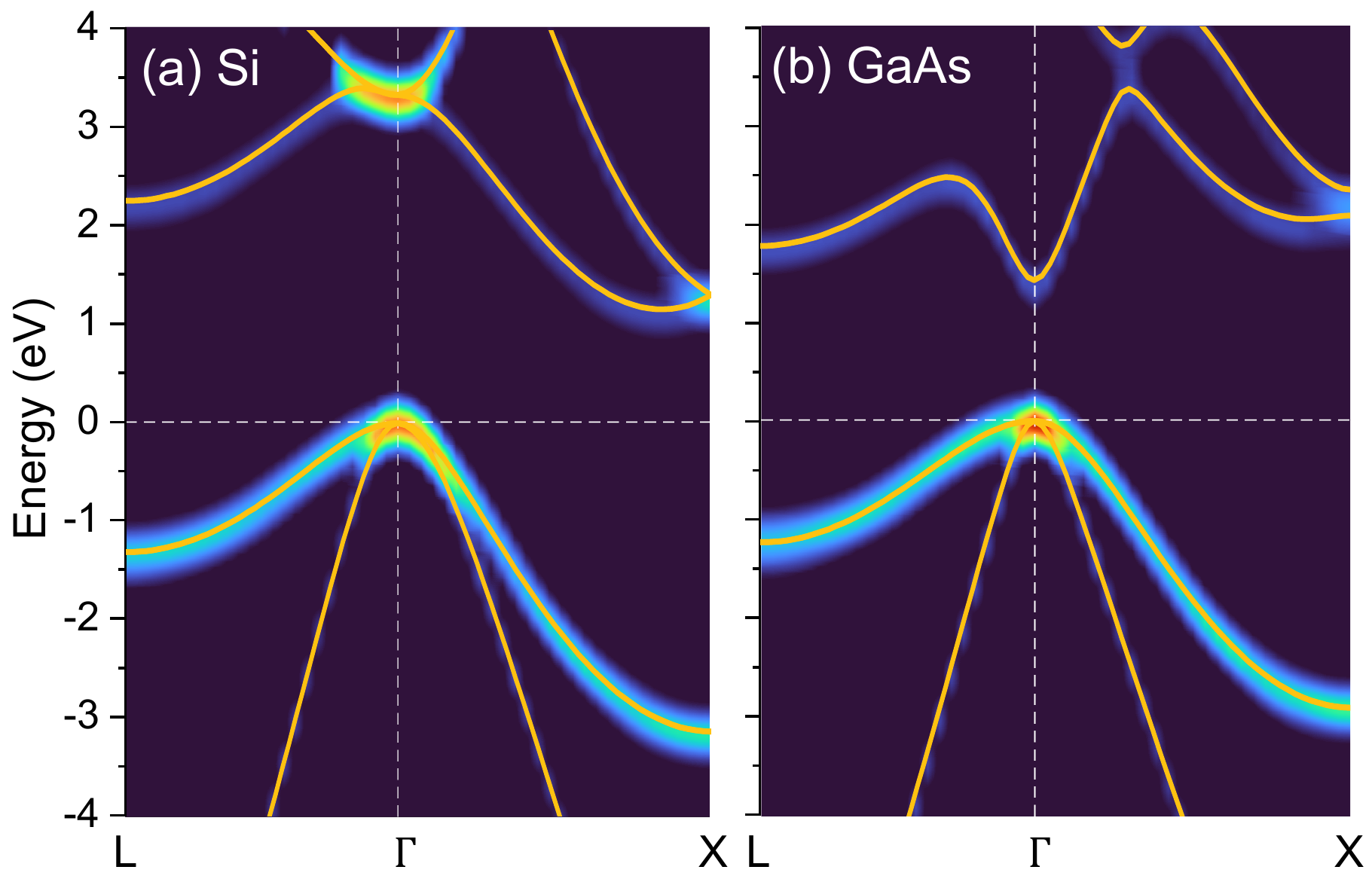}
\caption{
Unfolded band structures of (a) Si and (b) GaAs at room temperature in comparison with the corresponding band structures at $T = 0$~K (orange solid lines).
}
\label{fig:SiGaAs}
\end{figure}

Finally, we show that the drastic enhancement of effective masses due to thermal fluctuations is a unique feature in strongly anharmonic materials such as halide perovskites, and is in clear contrast with the situation in more conventional semiconductors.
We employ the same methodology as used for CsPbBr$_3$ to compute the unfolded band structures of Si and GaAs at room temperature [see the colormaps in Fig.~\ref{fig:SiGaAs}(a,b)].
Their electronic band structures at $T = 0$~K are also computed using the HSE hybrid functional [with the same mixing parameter (default value: 0.25) used for evaluating the room-temperature band structure] for comparison [see the orange solid lines in Fig.~\ref{fig:SiGaAs}(a,b)].

It is evident from Fig.~\ref{fig:SiGaAs}(a,b) that the smearing in the room-temperature band structures of both Si and GaAs is much weaker than that in CsPbBr$_3$, which is due to much less pronounced thermal fluctuations in Si and GaAs.
The dispersions of the band edges in the room-temperature band structures of Si and GaAs are nearly identical to those at $T=0$~K.
Quantitatively, the direction-averaged heavy-hole effective masses for both Si and GaAs vary by only around 5\% when changing the temperature from 0 to 300~K.
This is distinctly different from the behavior observed in CsPbBr$_3$.

\begin{figure}[ht]	
\includegraphics[width=86mm]{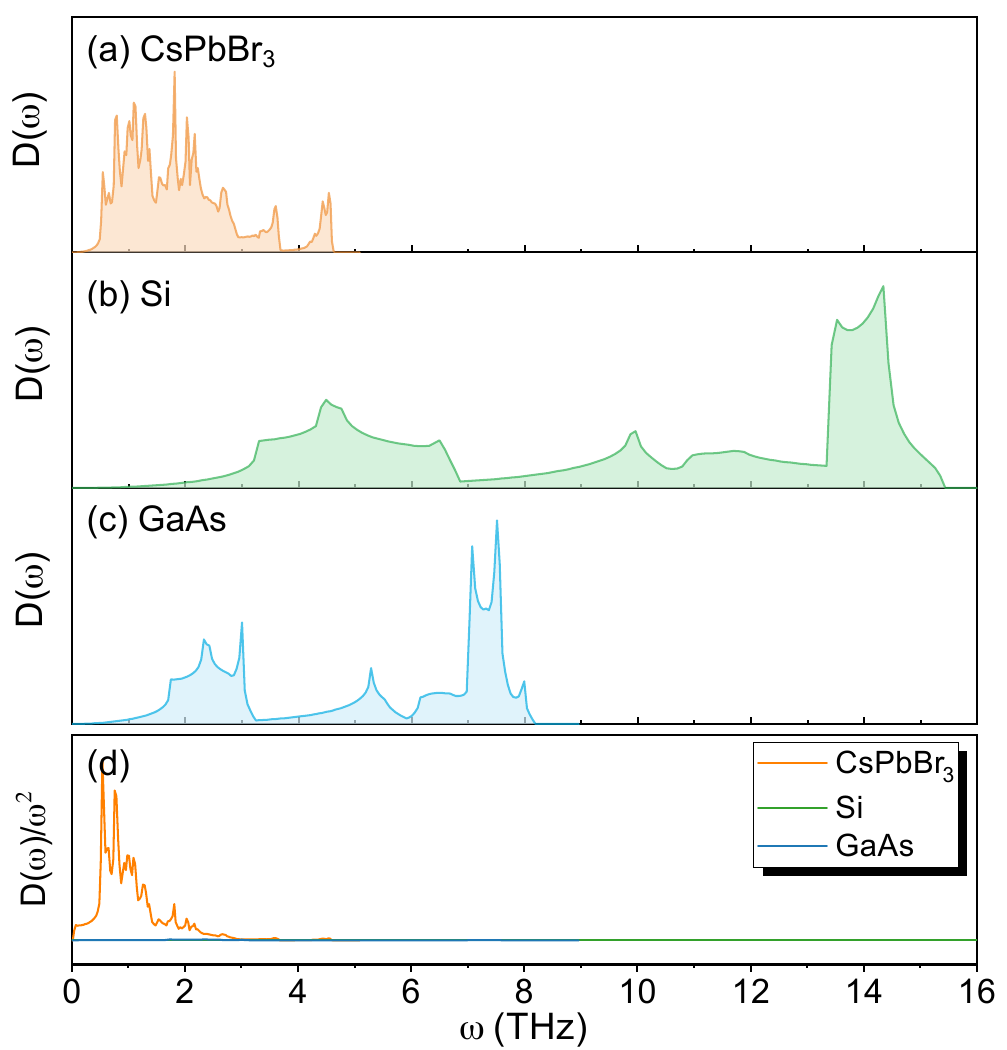}
\caption{
(a-c) Phonon density of states [$D(\omega)$] of (a) CsPbBr$_3$, (b) Si, and (c) GaAs.
(d) $D(\omega)/\omega^2$ as a function of $\omega$ for CsPbBr$_3$, Si, and GaAs.
}
\label{fig:phonon}
\end{figure}

The different levels of thermal fluctuations can be understood by analyzing the phonon density of states [$D(\omega)$] of the three materials in Fig.~\ref{fig:phonon}(a-c). 
The vibrational frequencies in CsPbBr$_3$ are drastically lower than those in Si and GaAs, making its vibrational modes much easier to be activated at low temperatures.
We also plot $D(\omega)/\omega^2$ as a function of $\omega$ in Fig.~\ref{fig:phonon}(d).
According to the Debye model for vibrational frequencies, $D(\omega)$ has a quadratic dependence on $\omega$ at low frequencies for normal harmonic crystals~\cite{buchenau_anharmonic_2003}.
The fact that CsPbBr$_3$ exhibits a distinct peak (commonly referred to as a boson peak~\cite{baggioli_universal_2019}) in the low-frequency region clearly indicates the existence of higher-order contributions.
The strong anharmonicity in CsPbBr$_3$ flattens the potential energy surface of lattice vibrations, enhancing the thermal fluctuations and disordering in CsPbBr$_3$ at room temperature.

In conclusion, we have shown that thermal lattice fluctuations significantly renormalize the electronic band structure of CsPbBr$_3$, which can well explain the previously observed hole effective mass in ARPES experiments.
Hence, there is no evidence for polaron formation from the ARPES experiments.
The strong thermal fluctuations in CsPbBr$_3$ are intimately related to its soft lattice with low vibrational frequencies and pronounced anharmonicity, in clear contrast with conventional semiconductors such as Si and GaAs.
Our study clarifies a controversy on the evidence of polaron formation in halide perovskites and demonstrates how to rigorously assess the finite-temperature band structure of strongly anharmonic materials.

\begin{acknowledgments}
B.Z., X.C., C.L., J.L., and X.Z. were supported by the National Natural Science Foundation of China (Grant No. 52172136).
C.G.VdW. was supported by the U.S. Department of Energy (DOE), Office of Science, Basic Energy Sciences (BES) under Award No.\ DE-SC0010689.
This research used resources of the National Energy Research Scientific Computing Center, a DOE Office of Science User Facility supported by the Office of Science of the U.S. Department of Energy under Contract No. DE-AC02-05CH11231 using NERSC award BES-ERCAP0021024.
The authors also acknowledge the Beijing Super Cloud Center (BSCC) for providing HPC resources that have contributed to the research results reported within this paper. URL: http://www.blsc.cn/
\end{acknowledgments}

\bibliography{bibliography}

\end{document}